\documentclass[prl,aps,twocolumn,showpacs]{revtex4-1}
\usepackage{amsmath}
\usepackage{amsfonts}
\usepackage{amssymb}
\usepackage{graphicx,hyperref}
\usepackage{mathptmx}

\newcommand{\be}{\begin {equation}}
\newcommand{\ee}{\end {equation}}
\newcommand{\beq}{\begin {eqnarray}}
\newcommand{\eeq}{\end {eqnarray}}

\hypersetup{
    colorlinks,    citecolor=black,    filecolor=black,    linkcolor=black,    urlcolor=black
}

\begin{document}

\title{Particle-in-Cell simulations of electron spin effects in plasmas}

\author{Gert Brodin}
\email[E-mail address: ]{gert.brodin@physics.umu.se}
\affiliation{Department of Physics, Ume{\aa} University, SE--901 87
Ume{\aa}, Sweden}

\author{Amol Holkundkar}
\email[E-mail address: ]{amol.holkundkar@physics.umu.se}
\altaffiliation[Currently at: ]{Department of Physics, Birla Institute of Technology and Science, Pilani, Rajasthan 333 031, India.
(amol.holkundkar@bits-pilani.ac.in)}
\affiliation{Department of Physics, Ume{\aa} University, SE--901 87
Ume{\aa}, Sweden}

\author{Mattias Marklund}
\email[E-mail address: ]{mattias.marklund@physics.umu.se}
\altaffiliation[Also at: ]{Department of Applied Physics, Chalmers University of Technology, SE--412 96 G\"oteborg, Sweden}
\affiliation{Department of Physics, Ume{\aa} University, SE--901 87
Ume{\aa}, Sweden}

\begin{abstract}
We have here developed a particle-in-cell code accounting for the magnetic
dipole force and for the magnetization currents associated with the electron
spin. The electrons is divided into spin-up and spin-down populations
relative to the magnetic field, where the magnetic dipole force acts in
opposite directions for the two species. To validate the code, we have
studied the wakefield generation by an electromagnetic pulse propagating
parallel to an external magnetic field. The properties of the generated
wakefield is shown to be in good quantitative agreement with previous
theoretical results. Generalizations of the code to account for more quantum
effects is discussed
\end{abstract}

\maketitle

\affiliation{Department of Physics, Ume{\aa} University, SE--901 87
Ume{\aa}, Sweden}



Recently much work has been devoted to the field of quantum plasmas. Many
aspects of the field are well described in books and review papers \cite%
{Bonitz -1998,Manfredi,glenzer-redmer,Shukla-Eliasson,Shukla-Eliasson-RMP}.
The interest in the field \ has been motivated by laboratory applications in
for example plasmonics \cite{Atwater-Plasmonics,Marklund-EPL-plasmonics},
quantum wells \cite{Manfredi-quantum-well} and high-density laser plasma
interaction \cite{glenzer-redmer,glenzer} as encountered e.g. in inertial
confinement fusion (ICF) schemes. Furthermore, astrophysical applications 
\cite{Astro} of quantum plasmas has played a role. The theoretical
approaches includes quantum kinetic theories such as e.g. the Kadanoff-Baym
kinetic \cite{Bonitz -1998} or the Wigner equation \cite{Manfredi}, various
quantum hydrodynamic equations \cite{Haas2005,Haas2010a,Haas2010b} and spin
kinetic theories \cite{Nature-Shukla-2009,Brodin2008,Zamanian2010,Asenjo2012}%
. Due to the complex nature of the governing equations, many problems
involving nonlinearities and/or inhomogeneities must be addressed
numerically. A successful method applied to \textit{classical plasmas} is
the celebrated particle-in-cell (PIC) approach (see e.g. \cite%
{Dawson-RMP,PIC-1996,PIC-2002,lpic++}). Due to the classicality of the
concepts this have had a limited impact to quantum plasmas. It should be
noted, however, that the Feynman path integral formulation has been used to
develop a PIC treatment \cite{QPIC-1,QPIC-2} that includes particle
dispersive effects in a semi-classical fashion. This is computationally
costly, and the number of quantum particles that can be included in the code
is far less than in the classical case \cite{QPIC-1,QPIC-2}.

In our paper we have chosen another approach in order to develop a PIC-code
including certain quantum features. We first note that the Wigner equation
reduces to the classical Vlasov equation for macroscopic scale lengths
longer than the thermal de Broglie wavelength. Hence particle dispersive
effects \cite{Manfredi,SW-19} can be neglected on such scales. On the other
hand, the physics associated with the electron spin (e.g. the magnetic
dipole force and the spin magnetization currents \cite%
{Brodin2008,Zamanian2010,Asenjo2012}) does not vanish for long scale
lengths. Hence, in this paper we will aim to develop a PIC code applicable
on macroscopic scales, accounting for the magnetic dipole force and the spin
magnetization currents. While a classical magnetic dipole moment fits very
well into the PIC-concept, it is clear that the difference between a
classical magnetic dipole moment and the spin must be acknowledged. The
present approach builds on the findings of Ref. \cite{NJP-spin}. It was then
observed that more elaborate models for the spin physics reduces to a simple
one for frequencies below the spin precession frequency (which is
approximately the same as the cyclotron frequency). Specifically this meant
that for a dynamical time scale slower than the cyclotron frequency, the
electrons could be modelled as consisting of two fluids, one with a spin up
state relative the magnetic field and one with a spin down state. The two
electron species are then subject to magnetic dipole forces acting in
opposite directions. A 1D PIC-code based on this concept has been developed,
and is tested in the present paper. In particular we let a short
electromagnetic (EM) pulse propagate along an external magnetic field in a
plasma, and study the wake field induced. Comparing with theoretical
results, we find that the PIC-code is able to reproduce previous findings 
\cite{Spin-pond-1,Misra-2010}.

Due to the classical nature of the PIC-concept, it is theoretically
difficult to generalize it to the quantum regime. Although the basic effect
of wave particle dispersion (as described by the Schroedinger equation) can
be handled to a certain extent (see e.g. Refs. \cite{QPIC-1,QPIC-2}) it is
computationally costly. In particular the number of particles needed to
capture particle dispersion is increased by a factor of at least a thousand.
On the other hand, the effect of particle dispersion is not important for
scale lengths much longer than the characteristic de Broglie wavelength \cite%
{Manfredi,Zamanian2010}. However, other quantum effects may play a role for
the long scale dynamics \cite{Brodin2008,Zamanian2010,Asenjo2012}. In
particular, when $\mu _{B}B/k_{B}T$ or $\mu _{B}B/m_{i}c_{A}^{2}$ approaches
unity (here $\mu _{B}=e\hbar /2m_{e}$ is the Bohr magneton, $B$ is a
characteristic magnetic field value, $T$ is a characteristic temperature, $e$
is the elementary charge, $\hbar $ is the Planck constant divided by $2\pi $%
, $m_{e}$ and $m_{i}$ is the electron and ion mass respectively, and $c_{A}$
is the characteristic Alfven velocity), the magnetic dipole force and the
magnetization current associated with the electron spin may become important 
\cite{Brodin2008, Lundin-PRE}. To a significant extent, the magnetic dipole
force associated with the spin is comparable to a classical magnetic dipole
force, and the magnetic moment associated with the magnetic field is
comparable to a classical magnetic moment. This means that spin effects can
be conceptually straightforward to include in a PIC-scheme. A complication
is that unlike a classical magnetic dipole moment, the probability
distribution of the spin for a single particle is by necessity spread out
(since the different spin components does not commute). This suggests that
spin-velocity correlations \cite{SV-Correlations} may affect the dynamics,
and that updating the variables using the Heisenberg equation of motion may
be somewhat inaccurate. However, the case where the dynamical time scale is
slower than the electron spin precession frequency (which is approximately
the same as the electron cyclotron frequency) and the spatial scale is
longer than the Larmor radius has been studied by Ref. \cite{NJP-spin}. It
was then deduced that the electrons can be described by a rather simple
model, where only two spin states, \ up and down relative to the magnetic
field are needed. The quantum effects then enter as a magnetic dipole force
in the equation of motion (in opposite directions for the two spin states,
which is described as two different species), and as a magnetization along
the magnetic field, which is proportional to the density difference between
the spin-up and spin-down electrons.

Based on the concepts described above, we will extend an 1D Particle-In-Cell
simulation (LPIC++) \ (see Ref. \cite{lpic++}) in order to incorporate the
effect of up- and down spin states in the governing dynamics. In this model
the ions are treated as a classical particles. However the electrons come in
two different species, one with spin up and the other with spin down. The
presence of an external magnetic field along the direction of laser
propagation is also included in the model. The governing equations for
particle motion with two electron species will now read as, 
\begin{equation}
m_{e}\frac{d\mathbf{v_{\downarrow }}}{dt}=q_{e}(\mathbf{E}+\mathbf{%
v_{\downarrow }}\times \mathbf{B})-\mu _{B}\nabla B
\end{equation}%
\begin{equation}
m_{e}\frac{d\mathbf{v_{\uparrow }}}{dt}=q_{e}(\mathbf{E}+\mathbf{v_{\uparrow
}}\times \mathbf{B})+\mu _{B}\nabla B
\end{equation}%
where the arrows indicate the different spin states and $B$ is the magnitude
of the magnetic field. Moreover Ampere's law is written as, 
\begin{equation}
\nabla \times (\mathbf{B}-\mu _{0}\mathbf{M})=\mu _{0}\mathbf{j}_{f}+\frac{1%
}{c^{2}}\frac{\partial \mathbf{E}}{\partial t}
\end{equation}%
where the free current including the ions is $\mathbf{j}_{f}=q_{e}(n_{%
\downarrow }\mathbf{v}_{\downarrow }+n_{\uparrow }\mathbf{v}_{\uparrow
})+q_{i}\mathbf{v}_{i}$, and the magnetization due to spin is $\mathbf{M}%
=\mu _{B}\mathbf{b}(n_{\downarrow }-n_{\uparrow })$, where $\mathbf{b}$ is
the unit vector in the direction of magnetic field, i.e. $\mathbf{b}=\mathbf{%
B}/B$. Two assumptions is necessary for this model to hold

i) That the spin-relaxation time is longer than the time scales of study.
Otherwise the two spin-states cannot be treated as independent fluids.

ii) That the time scale of study is longer than the spin precession time
scale. Otherwise the spin vector (and the magnetic moment) will not
necessarily point in the direction of the magnetic field.

\begin{figure}[t]
\centering \includegraphics[width=.8\columnwidth,angle=270]{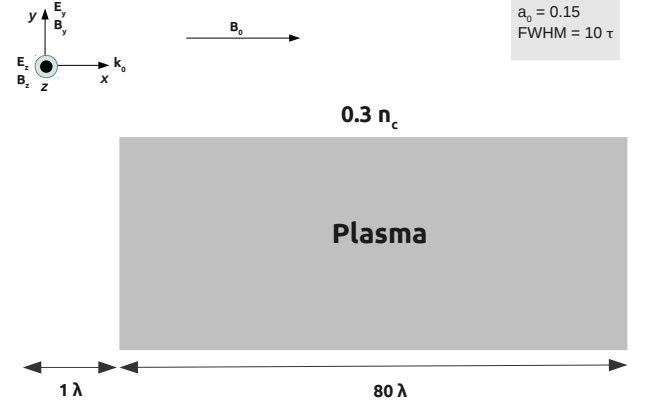}
\caption{Simulation geometry of the problem.}
\label{geometry}
\end{figure}

Before we start the studies it is convenient to rewrite the equations in
dimensionless form. We use the inverse laser frequency $\omega ^{-1}$ to
normalize time (i.e. the normalized time $t_{n}$ is $t_{n}=\omega t$), $%
c/\omega $ to normalize distance, the normalized spin-down (up) density is $%
n_{n\downarrow }\mathbf{_{(\uparrow )}}=\left( n_{\downarrow }\mathbf{%
_{(\uparrow )}}e^{2}/\varepsilon _{0}m\right) /\omega ^{2}$, and the
dimensionless electric and magnetic field are $\mathbf{E}_{n}=e\mathbf{E}%
/m_{e}\omega c$ and $\mathbf{B}_{n}=e\mathbf{B/}m_{e}\omega $, respectively.
Dividing the magnetic field into a homogeneous static field $B_{0}=B_{0}%
\widehat{\mathbf{x}}$ and a dynamic magnetic field, the basic equations in
normalized form becomes 
\begin{equation}
\frac{d\mathbf{v_{\downarrow }}}{dt_{n}}=\mathbf{E}+\mathbf{v_{\downarrow }}%
\times \left( \mathbf{B}-\frac{\omega _{c}}{\omega }\widehat{\mathbf{x}}%
\right) -\frac{\hbar \omega }{mc^{2}}\nabla B  \label{Norm-mom-1}
\end{equation}%
\begin{equation}
\frac{d\mathbf{v_{\uparrow }}}{dt_{n}}=\mathbf{E}+\mathbf{v_{\uparrow }}%
\times \left( \mathbf{B}-\frac{\omega _{c}}{\omega }\widehat{\mathbf{x}}%
\right) +\frac{\hbar \omega }{mc^{2}}\nabla B  \label{Norm-mom-2}
\end{equation}%
and 
\begin{equation}
\nabla \times \left( \mathbf{B}-\frac{\hbar \omega }{mc^{2}}\mathbf{b}%
(n_{\downarrow }-n_{\uparrow })\right) =(n_{\downarrow }\mathbf{%
v_{\downarrow }}+n_{\uparrow }\mathbf{v_{\uparrow }})+\frac{\partial \mathbf{%
E}}{\partial t}
\end{equation}%
Here for notational convenience the subscript $n$ denoting normalized
quantities have been omitted on all variables and operators. The term $%
(\omega _{c}/\omega )\mathbf{x}$, where $\omega _{c}=eB_{0}/m$ $\ $%
represents the constant static part of the magnetic field, and hence $%
\mathbf{B}$ is the dynamic part only. Consequently we have $B=\left\vert
(\omega _{c}/\omega )\mathbf{x+B}\right\vert $. \ For all simulations
presented below the dimensionless parameter $\hbar \omega /mc^{2}$ is chosen
to be $\hbar \omega /mc^{2}=2\pi \times 10^{-4}$. The typical simulation
geometry for the problem is shown in Fig. \ref{geometry}. For all the
results presented here the normalized electric field amplitude is considered
as $0.15$. The circularly polarized laser with FWHM duration of 10 period
times propagates through the plasma along the x direction. The external
magnetic field is considered along the x direction with amplitude in
dimensionless unit ($b_{0}=eB_{0}/m_{e}\omega $). With these parameters the
characteristic velocities of the problem (e.g. phase velocity and group
velocity) will be of the order of $c.$The thermal velocity is considered to
be much smaller than this, and hence the temperature is effectively zero in
our simulations. Finally, the plasma of length 80 wavelengths is taken to
have the total unperturbed electron density of $n_{\downarrow }+n_{\uparrow
}=0.3$ (Fig. \ref{geometry}).

\begin{figure}[t]
\centering \includegraphics[width=.7\columnwidth,angle=270]{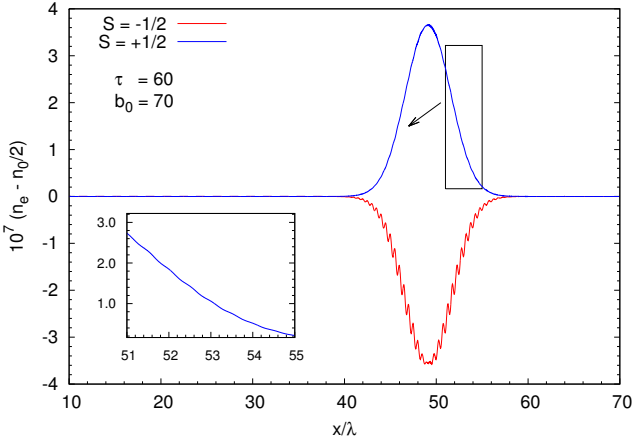}
\caption{Density perturbation with respect to unperturbed density for
electron species with spin $+1/2$ and $-1/2$ are presented at $60\protect%
\tau $ with external magnetic field $b_0 = 70$. Inset shows the magnified
version of the $+1/2$ spin state plot, where the presence of the small
second harmonic component can be seen. }
\label{den_per}
\end{figure}

\begin{figure}[b]
\centering \includegraphics[width=.7\columnwidth,angle=270]{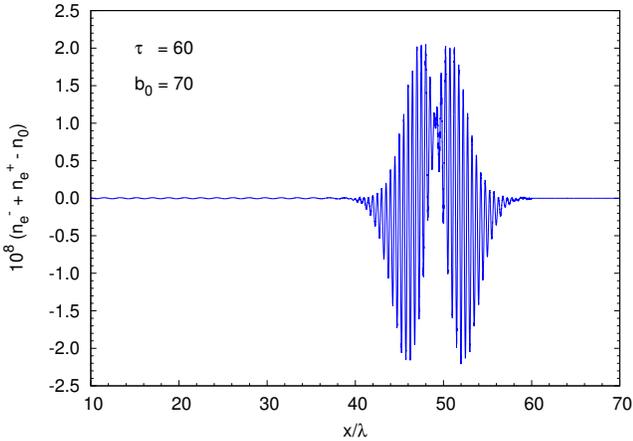}
\caption{Spatial profile of total electron density at $60\protect\tau$ for
the external magnetic field $b_0 = 70$. }
\label{den_add}
\end{figure}

In Fig. \ref{den_per} the density perturbation of the electrons with spin-up
and with spin-down is shown after the passage of a circularly polarized EM
wave. As can be seen the density perturbations of the two species of
electrons are almost opposite. This is also clear from Fig. \ref{den_add},
where the total (spin-up + spin-down) density perturbation is shown, which
is smaller than the individual density perturbations of the up- and down-
species by roughly a factor 20. Furthermore, we see that the total density
perturbation has a rapidly oscillating second harmonic component which is of
the same order of magnitude as the low-frequency component. Since the
electromagnetic wave is circularly polarized, we note that the second
harmonnic components are associated with the discrete particle (collisional)
effects of the PIC-code. To understand the results in more detail, we first
note that the spin properties in the given geometry only enters nonlinearly.
Since the magnetic field perturbation is purely transverse, we note that $%
\nabla B=\nabla \sqrt{B_{0}^{2}+B_{\bot }^{2}}\approx (1/2)\nabla (B_{\bot
}^{2}/B_{0})$, including only up to second order nonlinearities in the
transverse wave magnetic field $B_{\bot }$. The spin dependent part of the
perturbation is then induced by the magnetic dipole force, which is directed
in opposite directions for the spin-up and down species.

\begin{figure}[t]
\centering \includegraphics[width=.7\columnwidth,angle=270]{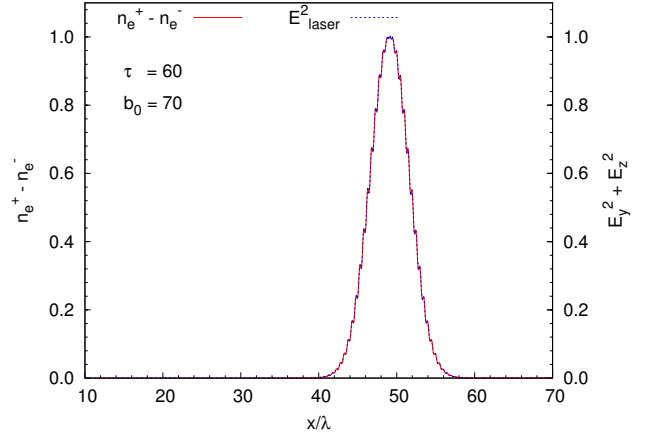}
\caption{Electron density response to the electric field of circularly
polarized laser pulse is presented. Both electric field and difference
between electron densities of both species are normalized to unity. It
should be noted that both curve are identical and it is not possible to
distinguish one from another in this figure.}
\label{driver}
\end{figure}

In order to verify the interpretation that the density difference between
the up- and down electron species comes is a direct response to the magnetic
dipole force of the wave field, we compare the profile of $\left\vert 
\mathbf{E}\right\vert ^{2}$(which is proportional to $B_{\bot }^{2}$) with
that of the density difference $n_{e}^{+}-n_{e}^{-}$. The result is shown in
Fig. \ref{driver}. It turns out that the matching is so accurate that two
separate curves cannot be recognized, as they are fully overlapping.

\begin{figure}[t]
\centering \includegraphics[width=.7\columnwidth,angle=270]{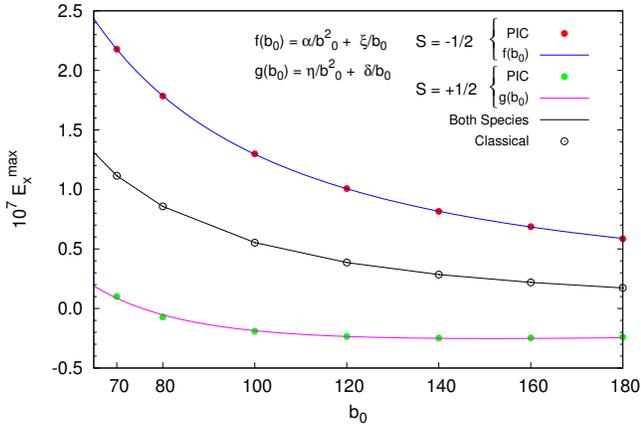}
\caption{Maximum amplitude of wakefield at $60\protect\tau$ is presented by
varying the magnitude of external magnetic for electrons species with
different spin states. For all the plasma electrons with spin down states
(red circles) the associated function $f(b_0)$ is fitted (blue line) with $%
\protect\alpha = 5.328\times 10^{-4}$ and $\protect\xi = 7.629\times 10^{-6}$%
. For all the plasma electrons with spin up states (green circles) the
associated function $g(b_0)$ is fitted (magenta line) with $\protect\eta =
5.753\times 10^{-4}$ and $\protect\delta = -7.615\times 10^{-6}$. For the
mixed electron species present in plasma (black line) nicely coincides with
the classical results (black open circles) with $\protect\mu_B = 0$.}
\label{max_ex}
\end{figure}

Next our aim is to make a quantitative comparison of the output from the
PIC-code with analytical results derived by Ref. \cite{Misra-2010}. Eqs.
(19) and (20) together with Eq. (20) of Ref. \cite{Misra-2010} show that for
strong magnetic fields ($\omega \ll \omega _{c}$) the peak electric
amplitude of the wakefield scales as 
\begin{equation*}
E_{\mathrm{peak}}\propto \frac{c_{1}(n_{0+}-n_{0-})}{B_{0}}+\frac{c_{2}}{%
B_{0}^{2}}
\end{equation*}%
{} where $c_{1}$ and $c_{2}$ are constants. Fitting the peak amplitude as a
function of the magnetic field as calculated from the PIC-code, with a curve
of this type, a good agreement is found, as shown in Fig. \ref{max_ex}.
Specifically three different cases are investigated. A pure spin up state, a
pure spin down state and a mixed state with $n_{0+}=n_{0-}.$The constant $%
c_{1}$ varies less than $0.2\%$ between the three cases, whereas $c_{2}$
varies slightly from $5.735\times 10^{-4}$ to $5.328\times 10^{-4}$ when
switching from a pure spin-up plasma to a spin down plasma.

\begin{figure}[h]
\centering \includegraphics[width=.7\columnwidth,angle=270]{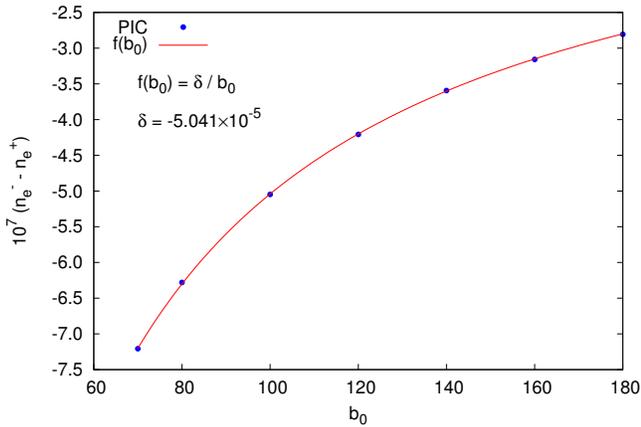}
\caption{The magnetization as a function of applied magnetic field (blue
circles) is fitted with function $f(b_{0})$ (red line).}
\label{magzatn}
\end{figure}

Another interesting variable to study is the spin polarization induced by
the EM wave for an plasma with equal number of particles in the spin up and
the spin down states initially. The spin polarization is monitored by
plotting the peak of the density difference $n_{+}-n_{-}$, which is also
closely related to the spin induced magnetization. According to the theory
(see e.g. Ref. \cite{Misra-2010}) the scaling with the magnetic field for $%
\omega \ll \omega _{c}$ is 
\begin{equation*}
n_{+}-n_{-}\propto B_{0}^{-1}
\end{equation*}%
which is confirmed in Fig. \ref{magzatn} to a very good accuracy.\ 

Extensions of PIC-codes to include the spin effects can be particularly
relevant for ICF plasmas. As demonstrated by Ref. \cite{Spin-pond-1}, the
spin dependent part of the ponderomotive force can give raise to significant
spin polarization for plasmas with metallic densities and higher, provided
the pulse length is not too short (cf Eq. (24) of Ref. \cite{Spin-pond-1})
Furthermore, the generation of strong quasi-static magnetic field in
laser-plasma interaction, even approaching giga-gauss values \cite{gigagauss}%
, means that the background plasma can be significantly spin-polarized,
since $\mu _{B}B/k_{B}T$ can approach unity in the fast ignition scenario
where the compression takes place before the heating \cite{Fast ignition}.
In the present study we have focused on validating validate the code against
the theoretical results of Ref. \cite{Misra-2010}, and hence picked
parameter values corresponding to astrophysical regimes. By comparing with
analytical theory we have verified that the code executes properly and with
good accuracy. In particular, studying the effects of a weakly modulated
high-frequency pulse, it has been shown that the low-frequency wake-field is
modified by the magnetic dipole force as predicted by theory (in particular,
see Fig. \ref{max_ex}). Furthermore, the spin polarization induced by the
high-frequency wave in an system that initially is not spin-polarized agrees
well with theory (see Fig. 6). This paper should be considered as a first
step towards a full spin PIC-code. Besides generalizing the code to higher
dimensions, it would be of much interest to include the spin dynamics (e.g.
spin precession), in which case it would be possible to resolve the dynamics
on the high-frequency cyclotron scale.

\end{document}